\documentclass[a4paper,10pt,twoside]{cpc-hepnp}

\usepackage{multicol}
\usepackage{graphicx}
\usepackage{booktabs}
\usepackage{amssymb,bm,mathrsfs,bbm,amscd}
\usepackage[tbtags]{amsmath}
\usepackage{lastpage}

\begin{document}

\fancyhead[co]{\footnotesize S. N. Chen and J. L. Ping: $X(1835)$, $X(2120)$, $X(2370)$ and $\eta(1760)$ in chiral quark model}


\title{$X(1835)$, $X(2120)$, $X(2370)$ and $\eta(1760)$ in chiral quark model\thanks{Supported by National Natural Science Foundation of China (11035006 and 11175088) }}

\author{Shaona Chen$^{1)}$ \quad Jialun Ping$^{1,2;1)}$\email{jlping@njnu.edu.cn (Corresponding author)}}

\maketitle

\address{%
$^1$ School of Physical Science and Technology, Nanjing Normal University, Nanjing
210097, China \\
$^2$ School of Physics and Microelectronics, Shandong University, Jinan 250100, China
}

\begin{abstract}
By analyzing the meson spectra obtained in the constituent quark
model, we find that the pseudoscalar mesons $\eta'(2^1S_0)$ ,
$\eta(4^1S_0)$, $\eta'(3^1S_0)$ and $\eta'(4^1S_0)$ are the possible
candidates of $\eta(1760)$, $X(1835)$, $X(2120)$ and $X(2370)$.
The strong decay widths of these pseudoscalars to all
the possible decay modes are calculated within the framework of the
$^3P_0$ model. Although the total width of $\eta'(2^1S_0)$ is compatible
with the experimental value of BES for $\eta(1760)$, the partial decay
width to $\omega\omega$ is too small, which is not consistent with
the result of BES. If the state $X(1835)$ is interpreted as
$\eta(4^1S_0)$, the total decay width is compatible with the
experimental data, and the main decay modes will be $\pi a_0(980)$
and $\pi a_0(1450)$, which needs to be checked. The assignment of
of $X(2120)$ and $X(2370)$ to $\eta'(3^1S_0)$ and $\eta'(4^1S_0)$
is also disfavored in the present calculation because of the
incompatibility of the decay widths.
\end{abstract}



\begin{multicols}{2}

\section{Introduction}

In 2005, the BES Collaboration observed a narrow peak in the
$\eta'\pi^+\pi^-$ invariant mass spectrum in the process
$J/\psi \rightarrow \eta'\pi^+\pi^-$ with a statistic significance of
7.7$\sigma$. Fitting with Breit-Wigner function yields mass and
width~\cite{BESII}
\begin{eqnarray*}
M & = & 1833.7 \pm 6.1(stat) \pm 2.7(syst) ~\mbox{MeV/c}^2 \\
\Gamma & = & 67.7 \pm 20.3(stat) \pm 7.7(syst) ~\mbox{MeV/c}^2,
\end{eqnarray*}
and the product branching fraction
\begin{eqnarray*}
& & B(J/\psi\rightarrow \gamma X(1835))
 B(X(1835)\rightarrow \pi^+\pi^-\eta') \\
 & & ~~~~~~~~~~= (2.2\pm 0.4(stat) \pm 0.4(syst)) \times 10^{-4}.
\end{eqnarray*}
BES-III confirmed it in the same process with statistical
significance larger than 20$\sigma$. The fitted mass and width are
$M=1836.5 \pm$3.0(stat)$^{+5.6}_{-2.1}$(syst) MeV/c$^2$,
$\Gamma$=190 $\pm$9(stat)$^{+38}_{-36}$(syst) MeV/c$^2$.
Meanwhile, another two new resonances, $X(2120)$ and $X(2370)$,
are also observed in the same process with the statistical significance
larger than 7.2$\sigma$ and 6.4$\sigma$, respectively. The fitted masses
and widths are~\cite{BESIII}
\begin{eqnarray*}
 M & = & 2122.4\pm 6.7(stat)^{+4.7}_{-2.7}(syst) ~\mbox{MeV/c}^2, \\
 M & = & 2376.3\pm 8.7(stat)^{+3.2}_{-4.3}(syst) ~\mbox{MeV/c}^2,
\end{eqnarray*}
and
\begin{eqnarray*}
\Gamma=83\pm 16(stat)^{+31}_{-17}(syst) ~\mbox{MeV/c}^2, \\
\Gamma=83\pm 17(stat)^{+44}_{-6}(syst) ~\mbox{MeV/c}^2,
\end{eqnarray*}
respectively. $\eta(1760)$, which its nature is in controversial,
was first reported by Mark III collaboration in the $J/\psi$
radiative decays to $\omega\omega$~\cite{MARKIII1} and
$\rho\rho$~\cite{MARKIII2}. And DM2 collaboration observed a large
bump peaked at 1.77 GeV/c$^2$ in $\omega\omega$ invariant mass
distribution in the process of
$J/\psi\rightarrow\gamma\omega\omega~(\omega\rightarrow\pi^+\pi^-\pi^0)$
\cite{DM21} and the study of the decays
$J/\psi\rightarrow\gamma\pi^+\pi^-\pi^+\pi^-$ and
$J/\psi\rightarrow\gamma\pi^+\pi^-\pi^0\pi^0$ showed that both decays
have a large $\rho\rho$ dynamics~\cite{DM22}. The fitted mass and width
are $M=1760\pm$11 MeV, $\Gamma=60\pm16$ MeV.
Recently BES collaboration reported its results on the decays
$J/\psi\rightarrow\gamma\omega\omega$,
$\omega\rightarrow\pi^+\pi^-\pi_0$~\cite{BESII2}. The mass and width
turn to be
$M=1744 \pm10$(stat)$\pm$15 MeV, $\Gamma$=$244^{+24}_{-21}\pm25$ Mev.

Many works have been devoted to the underlying structures of
$X(1835)$ and $\eta(1760)$~\cite{Klempt}. For $X(1835)$, Some interpret it as a
$p\bar{p}$ bound state~\cite{Ding,Dedonder,Wang,Liu}. By
calculating the mesonic decays of a baryonium resonance,
Ding {\em et al.} claimed that the $p\bar{p}$ bound state favors the
decay channel $X\rightarrow \eta4\pi$ over
$X\rightarrow \eta3\pi$~\cite{Ding}. In fact, it is just this work
that stimulates the observation of $J/\psi \rightarrow \eta'\pi^+\pi^-$
process in BES experiments. Using a semi-phenomenological potential model
that can describe all the $N\bar{N}$ scattering data, Dedonder {\em et al.}
found a broad spin-isospin singlets, $S$-wave quasi-bound state of
$N\bar{N}$, which can be used to explain the observed peak by
BES~\cite{Dedonder}. Z. G. Wang and S. L. Wang also calculated the mass
of $X(1835)$, which as a baryonium in the framework of QCD sum rule and
obtained a consistent result with experimental data. The large-$N_c$ QCD
is also applied to study the state $X(1835)$ as a baryonium~\cite{Liu}.
Interpretation of $X(1835)$ as a glueball or a glueball mixed with
pseudoscalar meson or baryonium is also proposed by using QCD sum
rule~\cite{Kochelev,He,Li,Hao}. Apart from these explanation for $X(1835)$
as an exotic state, the conventional $q\bar{q}$ picture of $X(1835)$ is
also proposed. Huang and Zhu studied the behavior of $X(1835)$ and thought
that it can be taken as the second radial excitation of $\eta^\prime(958)$,
in the effective Lagrangian approach~\cite{Huang}. The two-body decays of
$X(1835)$ as $3^1S_0$ are also calculated by quark-pair creation
(QPC, or $^3P_0$) model~\cite{PRD77}, the results show that the decay width
is sensitive to the mixing angle of two states $X_n=(u\bar{u}+d\bar{d})/\sqrt(2)$
and $X_s=s\bar{s}$. Recently J. S. Yu {\em et al.}
systematically studied the two-body strong decays and double pion decays of
$\eta$-family and assigned the $X(1835)$ to the second radial excitation of
$\eta^\prime(958)$, $X(2120$ and $X(2370)$ to the third and fourth radial
excitation of $\eta(548)/\eta^\prime(958)$, respectively~\cite{PRD83}.
For $\eta(1760)$, J. Vijande {\em et al.} assigned it to be $2^1S_0$
state of $s\bar{s}$ in the chiral quark model~\cite{JPG31}. The assignment
of $\eta(1760)$ to the second radial excitation of $\eta(548)$ is also
proposed by J. S. Yu {\em et al.}\cite{PRD83}. Li and Page
suggested it to be a gluonic meson~\cite{LiPage}. Glueball mixed with
$q\bar{q}$ picture of $\eta(1760)$ was also suggested by N. Wu {\em et al.}
\cite{Wu}. Stimulated by these experimental and theoretical work,
we shall study whether $\eta(1760)$, $X(1835)$, $X(2120)$ and $X(2370)$
can be described in the simplest system-$q\bar{q}$ system.

In this work, the pseudoscalar meson spectrum is determined by
the chiral quark model, the mixing angle between $X_n$ and $X_s$
is fixed through the system dynamics. Based on the mass spectrum,
the possible candidates of $X(1835)$, $X(2120)$, $X(2370)$ and
$\eta(1760)$ are assigned. Then the strong decay widths
of the states are calculated in the framework of $^3P_0$ model.
to see the assignment is reasonable or not.
The paper is organized as
follows: a brief review of $^3P_0$ model is given in
section 2. The chiral quark model is introduced and meson spectrum
and wave function scale parameter $\beta$ of the involved
mesons are obtained in section 3. The numerical result of the
strong decay are shown in section 4. The last section is a
summary.

\section{Review of $^3P_0$ model of meson decay }
The $^3P_0$ model also known as the Quark-Pair Creation (QPC) model,
applied to the decay of meson A to meson $B+C$ was first proposed by
Micu~\cite{Micu}, and then developed by Le Yaouanc, Ackleh, Roberts
{\rm et al}~\cite{Yaouanc1,Roberts,Ackleh}. The $^3P_0$ model assumes
that there is a pair of quark and antiquark created in
vacuum. The quantum number of the pair of quark and antiquark is
$J^{pc}=0^{++}$. Since vacuum is colorless and flavorless, so color
and flavor singlet should be satisfied. The created pair recombines
with the quark-antiquark pair in initial meson and form two mesons in
the final state in two possible ways, which is shown in Fig.1.
\begin{center}
\includegraphics[width=7cm]{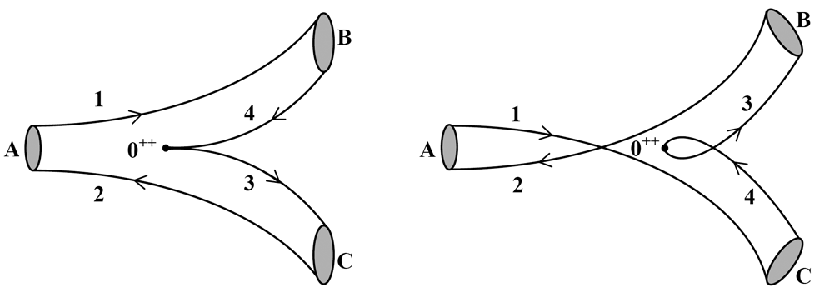}
\figcaption{\label{fig1} The two possible diagrams contributing to
$A\rightarrow B+C$ in the $^3P_0$ model.}
\end{center}

In the non-relativistic limit, the transition operator T takes form
as
\begin{eqnarray}
&&T=-3~\gamma\sum_m\langle 1m1-m|00\rangle\int
d\mathbf{p}_3d\mathbf{p}_4\delta^3(\mathbf{p}_3+\mathbf{p}_4)\nonumber\\
&&~~~~\times{\cal{Y}}^m_1(\frac{\mathbf{p}_3-\mathbf{p}_4}{2})
\chi^{34}_{1-m}\phi^{34}_0\omega^{34}_0b^\dagger_3(\mathbf{p}_3)
d^\dagger_4(\mathbf{p}_4),
\end{eqnarray}
where $\gamma$, which is a dimensionless parameter, represents the
strength of the quark-antiquark pair creation from the vacuum and can be
obtained  by fitting the experimental data. $\mathbf{p_3}$ and
$\mathbf{p_4}$ denote the momenta of the created quark and antiquark
respectively. $
{\cal{Y}}^m_l(\mathbf{p})=|p|^lY^m_l(\theta_p,\phi_p)$ is the $l$-th
solid harmonic polynomial that gives the momentum-space distribution
of the created quark-antiquark pair. $\chi^{34}_{1-m}$ reflects
triplet state of spin.
$\phi_0^{34}=(u\bar{u}+d\bar{d}+s\bar{s})/\sqrt{3}$ and
$\omega_0^{34}=(r\bar{r}+g\bar{g}+b\bar{b})/\sqrt{3}$ correspond to
flavor and color singlets, respectively.
$b^\dagger_3(\mathbf{p}_3)d^\dagger_4(\mathbf{p}_4)$ are the
creation operators of the quark and antiquark, respectively.

To depict the meson state, we define
\begin{eqnarray}
&& |A(n_A{}^{2S_A+1}L_{A}\,\mbox{}_{J_A M_{J_A}})(\mathbf{P}_A)\rangle\equiv\nonumber \\
&&\sqrt{2E_A}\sum_{M_{L_A},M_{S_A}}\langle L_A M_{L_A} S_A
M_{S_A}|J_AM_{J_A}\rangle \nonumber \\
&&\int
d\mathbf{p}_A\chi^{12}_{S_AM_{S_A}}\phi^{12}_A\omega^{12}_A
\left| q_1\left( \frac{m1}{m1+m2}\mathbf{P}_A+p_A\right)\right. \nonumber\\
&& \left. \bar{q}_2\left(\frac{m2}{m1+m2}\mathbf{P}_A+p_A\right)
 \right\rangle,
\end{eqnarray}
The wave function is normalized to
\begin{eqnarray}
& & \left\langle
A(n_A{}^{2S_A+1}L_{A}\,\mbox{}_{J_A M_{J_A}})(\mathbf{P}_A)
\right|  \\
&& \!\! \left. A(n_A{}^{2S_A+1}L_{A}\,\mbox{}_{J_A
M_{J_A}})(\mathbf{P}^{\prime}_A) \right\rangle
=2E_A\delta^3(\mathbf{P}_A-\mathbf{P}^{\prime}_A). \nonumber
\end{eqnarray}
where $ \chi^{12}_{S_AM_{S_A}},\phi^{12}_A,\omega^{12}_A$ represent
the spin, flavor and color wave function respectively;
$\mathbf{P_A}$ is the CM momentum of meson A, and
$\mathbf{p_A}=(m_2\mathbf{p_1}-m_1\mathbf{p_2})/(m_1+m_2)$ is the
relative momentum of $q\bar{q}$ pair. $n_A$ is the radial quantum number;
$|L_A,M_{L_A}\rangle,|S_A,M_{S_A}\rangle,|J_A,M_{J_A}\rangle$ are
the quantum number of orbit angular momentum between $q\bar{q}$ pair
in meson A, the total spin of the pair and the total angular
momentum, respectively; $\langle L_A M_{L_A} S_A M_{S_A}|J_A
M_{J_A}\rangle$ denotes a Clebsch-Gordan coefficient, $E_A$ is the
total energy of the meson.

To describe a strong decay process of $A\rightarrow B+C$, the
S-matrix is written out as
\begin{eqnarray}
\langle BC|S|A\rangle=I-2\pi i\delta(E_A-E_B-E_C)\langle
BC|T|A\rangle,
\end{eqnarray}
and then
\begin{eqnarray}
\langle
BC|T|A\rangle=\delta^3(\mathbf{P}_A-\mathbf{P}_B-\mathbf{P}_C)
{\cal{M}}^{M_{J_A}M_{J_B}M_{J_C}},
\end{eqnarray}
where ${\cal{M}}^{M_{J_A}M_{J_B}M_{J_C}}$ is the helicity amplitude
of $A\rightarrow B+C$. Taking the center of the mass frame of the
meson A: $\mathbf{P_A}=0$. One can obtained
${\cal{M}}^{M_{J_A}M_{J_B}M_{J_C}}$ for decay process in terms of
overlap integrals,

\end{multicols}
\ruleup
\begin{eqnarray}
{\cal{M}}^{M_{J_A}M_{J_B}M_{J_C}} & = & 3\gamma\sum_{\{M\}}
\langle L_AM_{L_A}S_AM_{S_A}|J_AM_{J_A}\rangle \langle
L_BM_{L_B}S_BM_{S_B}|J_BM_{J_B}\rangle \langle
L_CM_{L_C}S_CM_{S_C}|J_CM_{J_C}\rangle\nonumber\\
&\times&\langle1m1-m|00\rangle\langle\chi^{14}_{S_BM_{S_B}}\chi^{32}_{S_CM_{S_C}}
        |\chi^{12}_{S_AM_{S_A}}\chi^{34}_{1-m}\rangle[\langle\omega^{14}_B
        \omega^{32}_C|\omega^{12}_A\omega^{34}_0\rangle\langle
        \phi^{14}_B\phi^{32}_C|\phi^{12}_A\phi^{34}_0\rangle\nonumber\\
&\times&\mathcal{I}^{M_{L_A},m}_{M_{L_B},M_{L_C}}(\mathbf{P},m_1,m_2,m_3)
        +(-1)^{1+S_A+S_B+S_C}\langle\omega^{32}_B\omega^{14}_C|\omega^{12}_A\omega^{34}_0\rangle\langle
        \phi^{32}_B\phi^{14}_C|\phi^{12}_A\phi^{34}_0\rangle\nonumber\\
&\times&\mathcal{I}^{M_{L_A},m}_{M_{L_B},M_{L_C}}(-\mathbf{P},m_2,m_1,m_3)]
\end{eqnarray}
where $\{M\}=M_{L_A},M_{S_A},M_{L_B},M_{S_B},M_{L_C},M_{S_C},m$, The momentum space integral
$\mathcal{I}^{M_{L_A},m}_{M_{L_B},M_{L_C}}(\mathbf{P},m_1,m_2,m_3)$
and
$\mathcal{I}^{M_{L_A},m}_{M_{L_B},M_{L_C}}(\mathbf{-P},m_2,m_1,m_3)$
are given by
\begin{eqnarray}
\mathcal{I}^{M_{L_A},m}_{M_{L_B},M_{L_C}}(\mathbf{P},m_1,m_2,m_3)&=&\sqrt{8E_AE_BE_C}\int
d\mathbf{p}\,\mbox{}\psi^\ast_{n_BL_BM_{L_B}}
({\scriptstyle\frac{m_3}{m_1+m_3}}\mathbf{P}+\mathbf{p})
\psi^\ast_{n_CL_CM_{L_C}}
({\scriptstyle\frac{m_3}{m_2+m_3}}\mathbf{P}+\mathbf{p})\nonumber\\
&& \times \psi_{n_AL_AM_{L_A}}
(\mathbf{P}+\mathbf{p}){\cal{Y}}^m_1(\mathbf{p})
\end{eqnarray}
\begin{eqnarray}
\mathcal{I}^{M_{L_A},m}_{M_{L_B},M_{L_C}}(\mathbf{-P},m_2,m_1,m_3)&=&\sqrt{8E_AE_BE_C}\int
d\mathbf{p}\,\mbox{}\psi^\ast_{n_BL_BM_{L_B}}
({\scriptstyle-\frac{m_3}{m_1+m_3}}\mathbf{P}+\mathbf{p})
\psi^\ast_{n_CL_CM_{L_C}}
({\scriptstyle-\frac{m_3}{m_2+m_3}}\mathbf{P}+\mathbf{p})\nonumber\\
&&\times\psi_{n_AL_AM_{L_A}}
(-\mathbf{P}+\mathbf{p}){\cal{Y}}^m_1(\mathbf{p})
\end{eqnarray}

\ruledown
\vspace{2.0mm}

\begin{multicols}{2}

\noindent where
$\mathbf{P_B}=-\mathbf{P_C}=\mathbf{P},\mathbf{p}=\mathbf{p_3}$, and
$m_3$ is the mass of the created quark. The spacial wavefunction one
take is the simple harmonic oscillator (SHO) wavefunction. In
momentum-space, the SHO wavefunction reads
\begin{eqnarray}
&& \hspace{-1.0cm} \Psi_{nLM_L}(\mathbf{p})=(-1)^n(-i)^LR^{L+\frac{3}{2}}
\sqrt{\frac{2n!}{\Gamma(n+L+\frac{3}{2})}}\nonumber\\
&&~~\times\exp\left(-\frac{R^2p^2}{2}\right)L^{L+\frac{1}{2}}_n
\left(R^2p^2\right)\mathcal{Y}_{LM_L}(\mathbf{p}),
\label{showave}
\end{eqnarray}
here $\mathcal{Y}_{LM_L}(\mathbf{p})$ is the solid harmonic
polynomial; $R$ is the parameter of  SHO wavefunction; $\mathbf{p}$
is the relative momentum between $q\bar{q}$ pair within one meson;
$L^{L+\frac{1}{2}}_n\left(R^2p^2\right)$ is the Laguerre polynomial.

The decay width can be written as follows
\begin{eqnarray}
\Gamma = \pi^2
\frac{{|\textbf{P}|}}{M_A^2(1+\delta_{BC})}\sum_{JL}\Big
|\mathcal{M}^{J L}\Big|^2,\label{partialwidth}
\end{eqnarray}
where $\mathcal{M}^{J L}$ is the partial wave amplitude, which
is related to the helicity amplitude
${\cal M}^{M_{J_A}M_{J_B}M_{J_C}}$ via the Jacob-Wick
formula~\cite{Jacob}
\begin{eqnarray}
&& \hspace{-0.70cm} {\mathcal{M}}^{J L}(A\rightarrow BC)
= \frac{\sqrt{2 L+1}}{2 J_A+1}
\!\! \sum_{M_{J_B},M_{J_C}} \!\!\!\! \langle L 0 J M_{J_A}|J_A
M_{J_A}\rangle \nonumber \\
&& \!\! \times\langle J_B M_{J_B}
J_C M_{J_C} | J M_{J_A} \rangle \mathcal{M}^{M_{J_A} M_{J_B}
M_{J_C}}({\textbf{P}}),
\end{eqnarray}
where $\mathbf{J}=\mathbf{J}_B+\mathbf{J}_C$, $\mathbf{J}_{A}
=\mathbf{J}_{B}+\mathbf{J}_C+\mathbf{L}$.
Then the decay width in terms of the partial wave
amplitude is taken as,
where $|\textbf{P}|=|\textbf{P}_{B}|=|\textbf{P}_{C}|$. According to
the calculation of 2-body phase space, one can get
\[|\textbf{P}|=\frac{\sqrt{[M^2_A-(M_B+M_C)^2][M^2_A-(M_B-M_C)^2]}}{2M_A},\]
where $M_A$, $M_B$, and $M_C$ are the masses of the meson $A$, $B$,
and $C$, respectively.

\section{The masses of the mesons}

To calculate the meson spectrum, a QCD-inspired model, constituent
quark model, is used. The model incorporates the perturbative (one gluon
exchange) and nonperturbative (color confinement and spontaneous
breaking of chiral symmetry) properties of QCD. The constituent quark
mass originates from the spontaneous breaking of chiral symmetry and
consequently constituent quarks should interact through the exchange
of Goldstone bosons~\cite{manohar}, in addition to the one-gluon-exchange.
To describe the hadron-hadron interaction, the chiral partner of pion,
$\sigma$-meson, is also used. So the model Hamiltonian is
\begin{equation}
H = m_1+m_2+\frac{\mathbf{p}^2}{2\mu}+V^C+V^G+V^{\chi}+V^{\sigma},
\end{equation}
\begin{equation*}
V^C={\boldmath{\mbox{$\lambda$}}}^c_{1}\cdot {\boldmath{\mbox{$\lambda$}}}^c_{2}
  \left[-a_{c} \left( 1- e^{-\mu_c\,r} \right)+ \Delta \right]+V^C_{SO}
\end{equation*}
\begin{eqnarray}
V_{SO}^{C} &=&-{\boldmath{\mbox{$\lambda$}}}^c_{1}\cdot
{\boldmath{\mbox{$\lambda$}}}^c_{2}
\frac{a_c \mu_c e^{-\mu_c r}}{4m_{1}^{2}m_{2}^{2}\,r}
\left[ \left( m_{1}^{2}+m_{2}^{2} \right)(1-2a_{s}) \right. \nonumber \\
& & \left. +4m_{i}m_{j}(1-a_{s}) \right] ~\mathbf{S} \cdot \mathbf{L}
\nonumber \\
V^{G} &= & V^{G}_C+V^{G}_{SO}+V^{G}_{T} \nonumber \\
V_C^G & =& \frac{\alpha_s}{4} {\boldmath{\mbox{$\lambda$}}}^c_{1}
 \cdot {\boldmath{\mbox{$\lambda$}}}^c_{2}
 \left\{ \frac{1}{r}-\frac{{\boldmath{\mbox{$\sigma$}}}_{1}
 \cdot {\boldmath{\mbox{$\sigma$}}}_{2}}{6m_1 m_2}
  \frac{e^{-r/r_{0}(\mu)}}{r\, r_0^2(\mu)} \right\} \nonumber \\
V_{OGE}^{SO} & =&
-{\frac{\alpha_s}{16}}{\frac{{\boldmath{\mbox{$\lambda$}}}^c_{1}
 \cdot {\boldmath{\mbox{$\lambda$}}}^c_{2}}{{m_1^2 m_2^2}}}
  \left[\frac{1}{r^3} -
 \frac{e^{-r/r_g(\mu)}}{r^3}
 \left( 1 + \frac{r}{r_g(\mu)}  \right) \right] \nonumber \\
& &
\left[ \left( (m_1+m_2)^2+2m_1 m_2\right) \mathbf{S} \cdot \mathbf{L}
\right], \nonumber \\
V_{OGE}^{T} & =& -\frac{1}{16}\frac{\alpha_s}{m_1 m_2}
 {\boldmath{\mbox{$\lambda$}}}^c_{1}
 \cdot {\boldmath{\mbox{$\lambda$}}}^c_{2} \left[ \frac{1}{r^3}
 - \frac{e^{-r/r_g(\mu)}}{r} \right. \nonumber \\
 & & \left. \left( \frac{1}{r^2} +
 \frac{1}{3 r_g^2(\mu)} + \frac{1}{r\,r_g(\mu)} \right)
\right] S_{12} , \nonumber \\
V_{\chi} & = & \left(v_{\pi}^{C}+v_{\pi}^{T}\right)
 \sum_{a=1}^{3} \lambda_{1}^{a}~\lambda_{2}^{a}
 + \left(v_{K}^{C}+v_{K}^{T}\right)
 \sum_{a=4}^{7} \lambda_{1}^{a}~\lambda_{2}^{a}
  \nonumber \\
& & + \left(v_{\eta}^{C}+v_{\eta}^{T}\right)
\left( \lambda_{1}^{8}~\lambda_{2}^{8} \cos\theta_P
  -\lambda_{1}^{0}~\lambda_{2}^{0}\sin\theta_P \right)
  \nonumber \\
v_{\chi}^C & = & C_1 \left[ Y(m_{\chi}\,r)
 -\frac{\Lambda_{\chi}^3}{m_{\chi}^3}
 Y(\Lambda_{\chi}\,r)\right] {\boldmath{\mbox{$\sigma$}}}_{1}
 \cdot {\boldmath{\mbox{$\sigma$}}}_{2}, \nonumber \\
v_{\chi}^T & = & C_1 \left[ H(m_{\chi}\,r)
 -\frac{\Lambda_{\chi}^3}{m_{\chi}^3}
 H(\Lambda_{\chi}\,r)\right] S_{12} \nonumber \\
C_1 &=& \frac{g_{ch}^2}{4\pi} \frac{m_{\chi}^2}{12m_1 m_2}
 \frac{\Lambda_{\chi}^2}{\Lambda_{\chi}^2- m_{\chi}^2} m_{\chi},
 \quad \chi=\pi,K,\eta, \nonumber \\
V_{\sigma} & = & -C_2 \left[ Y(m_{\sigma}r)
 -\frac{\Lambda_{\sigma}}{m_{\sigma}}
 Y(\Lambda_{\sigma}r) \right]+V_{\sigma}^{SO}, \nonumber \\
V_{\sigma}^{SO} & = & -C_2 \frac{m_{\sigma}^2}{2m_1 m_2}
 \left[ G(m_{\sigma}r) -\frac{\Lambda_{\sigma}^3}{m_{\sigma}^3}
 G(\Lambda_{\sigma}r) \right] \mathbf{S} \cdot \mathbf{L}
 \nonumber \\
 C_2 &=& \frac{g_{ch}^2}{4\pi}
 \frac{\Lambda_{\sigma}^2}{\Lambda_{\sigma}^2-m_{\sigma}^2} m_{\sigma}
 \nonumber \\
 && \hspace{-1.0cm} S_{12}=3({\boldmath{\mbox{$\sigma$}}}_{1} \cdot
 \mathbf{r})({\boldmath{\mbox{$\sigma$}}}_{2}\cdot \mathbf{r})
 -{\boldmath{\mbox{$\sigma$}}}_{1}
 \cdot {\boldmath{\mbox{$\sigma$}}}_{2}, \nonumber
\end{eqnarray}
\begin{eqnarray}
 && \hspace{-1.0cm} Y(x)=\frac{e^{-x}}{x},
 \quad H(x)=\left(1+\frac{3}{x}+\frac{3}{x^2}\right)Y(x),
 \nonumber \\
 && \hspace{-1.0cm} G(x)=\left(1+\frac{1}{x}\right)
 \frac{Y(x)}{x}, \nonumber
\end{eqnarray}
where $r=|\mathbf{r}_1-\mathbf{r}_2|$ and
$\mathbf{p}=(\mathbf{p}_1-\mathbf{p}_2)/2$, $r_0(\mu)=\hat{r}_0/\mu$,
$r_g(\mu)=\hat{r}_g/\mu$. Other symbols have their
usual meanings. The effective running coupling constant
is given by
\begin{equation}
\alpha_s(\mu)=\frac{\alpha_0}{\ln\left({{\mu^2+\mu^2_0}
\over\Lambda_0^2}\right)}, \label{asf}
\end{equation}
where $\mu$ is the reduced mass of the $q\bar{q}$ system.
The chiral coupling constant $g_{ch}$
is determined from the $\pi NN$ coupling constant through
\begin{equation}
\frac{g_{ch}^2}{4\pi}=\left( \frac{3}{5}\right)^2
\frac{g_{\pi NN}^2}{4\pi}\frac{m_{u,d}^2}{m_{N}^2}.
\end{equation}
The meson spectrum is obtained by solving the
Schr\"{o}dinger equation,
\begin{eqnarray}
H\Psi & = & E\Psi, \\
\Psi &= & \left[ \psi_{nLM_L}\chi_{SM_S} \right]_{JM_J}\chi_c\chi_f,
\end{eqnarray}
\begin{center}
\tabcaption{Model parameters. The masses of mesons $\pi,K,\eta$ take the
 experimental values.\label{tab2}}
\footnotesize
\begin{tabular*}{80mm}{c@{\extracolsep{\fill}}cccccc}
\toprule
$~~m_{u,d}~~$ & $~~m_s~~$   & $a_c$   & $\mu_c$     & $\Delta$    & $a_s$ \\
     MeV      &    MeV      &  MeV    & fm$^{-1}$   &   MeV       &   -   \\ \hline
     313      &    555      &  430    &   0.7       & 181.10      & 0.777 \\ \hline
$\alpha_0$    & $\Lambda_0$ & $\mu_0$ & $\hat{r}_0$ & $\hat{r}_g$ &       \\
     -        & fm$^{-1}$   &  MeV    &  MeV fm     & MeV fm      &       \\ \hline
   2.118      & 0.113       &  36.976 &  28.170     & 34.500      &       \\ \hline
$\Lambda_{\pi}$ & $\Lambda_{\sigma}$ & $\Lambda_{K}$ & $\Lambda_{\eta}$ &
  $g_{ch}^2/4\pi$ & $\theta_P$  \\
  fm$^{-1}$   & fm$^{-1}$   &   fm$^{-1}$   & fm$^{-1}$  & -  &  $^o$   \\ \hline
 4.20 & 4.20 & 5.20 & 5.20 & 0.54   &  -15  \\ \bottomrule
\end{tabular*}
\end{center}

\noindent where $\chi_{SM_S},\chi_c,\chi_f$ are spin, color and flavor wavefunctions
of the meson, respectively and can be constructed through the symmetry.
The spatial wavefunction
$\psi_{nLM_{L}}=R_{nL}(r)Y_{LM_L}(\Omega)$ is obtained by solving
the second-order differential equation. The efficient numerical
method: Numerov method~\cite{Koonin} is used here.
The model parameters, which are listed in Table \ref{tab2}, are fixed by
fitting the experimental data of meson spectrum.
Parts of the obtained meson spectrum are shown in Tables \ref{tab3}
and \ref{tab4}.
The detailed results can be found in Ref.~\cite{JPG31}.
To calculate the strong decay of mesons analytically in $^3P_0$ model,
the obtained radial part of the spacial wavefunction $R_{nL}(r)$ is fitted
by the simple harmonic oscillator (SHO),
\begin{eqnarray}
R_{nL}(r) & =& \beta^{(L+\frac{3}{2})} \sqrt{\frac{2n!}
{\Gamma(n+L+\frac{3}{2})}} \exp\left(-\frac{\beta^{2}r^2}{2}\right)
\nonumber\\
&&  r^L L^{L+\frac{1}{2}}_n\left(\beta^{2}r^2\right).
\label{Rnl}
\end{eqnarray}
The fitted values of parameter $\beta$ are also listed in Table 3.
\begin{center}
\tabcaption{The mass of $I=1,\frac{1}{2}$ mesons and the values of fitted
  $\beta$.\label{tab3}}
\footnotesize
\begin{tabular*}{80mm}{c@{\extracolsep{\fill}}cccccc}
\toprule
$n^{2S+1}L_J$ & states  & Isospin  & Mass  & $\beta$ & $R$ \\
   &   &   &  (MeV)  & (fm$^{-1}$) & (GeV$^{-1}$) \\ \hline
$1^1S_0$ & $\pi$       & 1       & 139      & 2.308 & 2.196 \\
$2^1S_0$ & $\pi(1300)$ & 1       & 1288     & 1.434 & 3.534 \\
$1^3S_1$&$\rho$        & 1       & 772      & 1.438 & 3.522 \\
$2^3S_1$&$\rho$(1450)&1       &1478         &1.096  & 4.624 \\
$1^1P_1$&$b_1(1235)$ &1       &1234         &1.243  & 4.077 \\
$1^3P_0$&$a_0(980)$  &1       &984          &1.473  & 3.440 \\
$2^3P_0$&$a_0(1450)$ &1       &1587         &1.125  & 4.505 \\
$1^3P_1$&$a_1(1260)$ &1       &1205         &1.300  & 3.898 \\
$1^3P_2$&$a_2(1320)$ &1       &1327         &1.106  & 4.582 \\
$1^3P_2$&$a_2(1700)$ &1       &1732         &0.890  & 5.694 \\ \hline
$1^1S_0$& $K$           &1/2     &496       &2.313  & 2.191 \\
$2^1S_0$& $K(1460)$     &1/2     &1472      &1.545  & 3.280 \\
$1^3S_1$& $K^*(892)$    &1/2     &910       &1.629  & 3.111 \\
$2^3S_1$& $K(1630)$     &1/2     &1620      &1.262  & 4.016 \\
$1^1P_1$& $K_1(1400)$   &1/2     &1414      &1.371  & 3.696 \\
$1^3P_0$& $K_0^*(1430)$ &1/2    &1213       &1.572  & 3.224 \\
$2^3P_0$& $K_0^*(1950)$ &1/2    &1768       &1.243  & 4.077 \\
$1^3P_1$& $K_1(273)$    &1/2     &1352      &1.435  & 3.531 \\
$1^3P_2$& $K_2^*(1430)$ &1/2    &1450       &1.572  & 3.224 \\
$1^3D_1$& $K_1(1680)$   &1/2    &1698       &1.205  & 4.206 \\
\bottomrule
\end{tabular*}
\end{center}

For $I=0$ states, there are two types of them, one is composed of
$u,d$-quark and $\bar{u},\bar{d}$-antiquark, another is composed of
$s$-quark and $\bar{s}$-antiquark. They are mixed in the flavor SU(3)
symmetry to form flavor singlet and octet. However, flavor SU(3) is
broken. In experiments, we have $\eta$ and $\eta^{\prime}$ instead of
$\eta_1$ and $\eta_8$ for pseudoscalar. In the present calculation,
flavor SU(3) symmetry is not used, so we have flavor wavefunctions
$X_n$ and $X_s$. As a consequence of $K$-meson exchange, they are
mixed. To obtain the masses of $I=0$ states, the following procedure
is taken. First, solving the Schr\"{o}dinger equation for $X_n$
and $X_s$ separately ($K$-meson exchange is not employed). Secondly,
by using the wavefunctions $\Psi_n$ and $\Psi_s$ obtained in the
first step and taking into
account of $K$-meson exchange, the eigen-energies and eigen-states
can be obtained by diagonalizing the Hamiltonian matrix
\begin{eqnarray}
 \left(\begin{array}{cc}
             H_{nn}& H_{ns}\\
             H_{sn}& H_{ss}\\
            \end{array}\right)\left(\begin{array}{c}
             C_n\\
             C_s\\
            \end{array}\right)=E\left(\begin{array}{c}
             C_n\\
             C_s\\
            \end{array}\right).
\end{eqnarray}
where $H_{nn}=\langle \Psi_n |H|\Psi_n\rangle$,
$H_{ns}=\langle \Psi_n |V_K|\Psi_s\rangle=H_{sn}$ and
$H_{ss}=\langle \Psi_s |H|\Psi_s\rangle$. The eigen-state is
$|\Psi\rangle=C_n| \Psi_n\rangle+C_s|\Psi_s\rangle$.
The obtained eigen-energies and eigen-states are shown in
Table \ref{tab4}. From Table \ref{tab4}, one finds that
$\eta(1760),X(1835),X(2120),X(2370)$ may be interpreted
as $\eta'(2^1S_0),\eta(4^1S_0),\eta'(3^1S_0)$ and $\eta'(4^1S_0)$
respectively by comparing the theoretical masses with the
experimental data. To check these assignments, the decay properties of
the states should be studied, which is discussed in the next section.

\end{multicols}

\begin{center}
\tabcaption{The masses of $I=0$ mesons and the value of fitted $\beta$
($\beta=C_n^2\beta_n+C_s^2\beta_s$).\label{tab4}}
\footnotesize
\begin{tabular*}{170mm}{c@{\extracolsep{\fill}}cccccc}
\toprule
$(nL)J^{PC}$ & states & Mass (MeV) & $C_n$ & $C_s$ & $\beta$(fm$^{-1}$)
 & $R$(GeV$^{-1}$) \\ \hline
$1^1S_0$ & $\eta$       & 572    & $8.6564\times10^{-1}$ &
  $-5.0066\times 10^{-1}$ & 1.732693 & 2.924 \\
$1^1S_0$&$\eta'(958)$   &956    &$5.0066\times10^{-1}$   &$8.6564\times10^{-1}$ & 2.064307 & 2.455 \\
$2^1S_0$&$\eta(1295)$   &1290&$9.6360\times10^{-1}$ &
$-2.67323\times10^{-1}$ & 1.183-1.666 & 3.041-4.284\\
$2^1S_0$&$\eta'(1760)$  &1795   &$2.6732\times10^{-1}$   &$9.6360\times10^{-1}$ & 1.183-1.666 & 3.041-4.284 \\
$3^1S_0$&$\eta(3S)$     &1563&$9.9350\times10^{-1}$&
$-1.1380\times10^{-1}$ & 0.929-1.360 & 3.726-5.455 \\
$3^1S_0$&$\eta'(3S)$    &2276&$1.1380\times10^{-1}$&
$9.9350\times10^{-1}$ & 0.929-1.360 & 3.726-5.455 \\
$4^1S_0$&$\eta(4S)$     &1807&$9.9935\times10^{-1}$&
$-3.5928\times10^{-2}$ & 0.6725-1.0995 & 4.607-7.530 \\
$4^1S_0$&$\eta'(4S)$ &2390&$3.5928\times10^{-2}$ &
$9.9935\times10^{-1}$ & 0.6725-1.0995 & 4.607-7.530 \\\hline
$1^3S_1$&$\omega(782)$  &691 &$9.9499\times10^{-1}$&
 $9.9967\times10^{-2}$ & 1.547 & 3.276 \\
$1^3S_1$&$\phi(1020)$   &1020&$-9.9967\times10^{-2}$&
$9.9499\times10^{-1}$& 1.918 & 2.642 \\
$2^3S_1$&$\omega(1420)$ &1444&$9.9852\times10^{-1}$ &
$5.4331\times10^{-2}$ & 1.163 & 4.357 \\
$2^3S_1$&$\phi(1680)$  &1726&$-5.4331\times10^{-2}$&
$9.9852\times10^{-1}$ & 1.506 & 3.365 \\\hline
$1^1P_1$&$h_1(1170)$    &1257 &1.0&0& 1.202 & 4.216 \\
$1^1P_1$&$h_1'$        &1511 &0&1.0& 1.581 & 3.205 \\
$1^3P_2$&$f_2(1270)$    &1311&1.0&0   & 1.112 & 4.557 \\
$1^3P_2$ & $f_2'(1525)$   & 1556 & 0 & 1.0 & 1.496 & 3.387 \\
\bottomrule
\end{tabular*}
\end{center}

\begin{multicols}{2}

\section{The strong decay of the candidates for
$\eta(1760),X(1835),X(2120),X(2370)$}

$\eta$, $\eta^{\prime}$ and their radial excitations have the same
quantum numbers $IJ^{PC}=00^{-+}$. According to the $^3P_0$ model
discussed above, the isospins of mesons $B$ and $C$ can takes
the values $I=0,1/2$, or 1 with the condition
$\mathbf{I}_B+\mathbf{I}_C=\mathbf{I}_A$. If not forbidden
kinetically, the allowed decay modes of $\eta(\eta^{\prime})$
family are listed in Table \ref{tab1}.

\end{multicols}

\begin{center}
\tabcaption{Allowed decay modes and the amplitudes of the radial excited states
of $\eta$ and $\eta^{\prime}$.
 For $X_n$ decay, $\phi_f= \sqrt{\frac{1}{2}},~\sqrt{\frac{1}{3}},~\sqrt{\frac{1}{6}},~0$ for
 $I_B=I_C=1,~1/2,~0(X_n),~0(X_s)$ and for $X_s$ decay, $\phi_f=0,~\sqrt{\frac{2}{3}},~0,~
 \sqrt{\frac{1}{3}}$ for $I_B=I_C=1,~1/2,~0(X_n),~0(X_s)$.\label{tab1}}
\footnotesize
\begin{tabular*}{170mm}{c@{\extracolsep{\fill}}ccc}
\toprule
$X \rightarrow ^1S_0+^3P_0$ & $\pi a_0(980), ~\pi a_0(1450), ~\pi(1300)a_0(980),$
 & $ M^{JL}=M^{00}=M^{000}$ \\
 & $KK^*_0(1430),~KK^*_0(1950)$ & $M^{000}=\sqrt{\frac{1}{36}}
 (I^{-1,-1}_{0,0}+I^{0,0}_{0,0}+I^{1,1}_{0,0}) \phi_f$ \\ \hline
$X \rightarrow ^1S_0+ ^3P_2$ & $\pi a_2(1320),~\pi a_2(1700),~KK^*_2(1430),
 ~\eta f_2(1270),$ & $M^{JL}=M^{22}=M^{000}$ \\
 & $\eta^{\prime} f_2(1270), ~\eta f^{\prime}_2(1525)$ &
 $M^{000}=\sqrt{\frac{1}{72}}(I^{-1,-1}_{0,0}-2M^{0-11})\phi_f$ \\ \hline
$X \rightarrow ^1S_0+^3S_1$ & $K K^*,~KK^*(1410), ~K(1460)K^*$ & $M^{JL}=M^{11}=-M^{000}$ \\
  & & $M^{000}=-\sqrt{\frac{1}{12}}I^{0,0}_{0,0}\phi_f$ \\ \hline
$X \rightarrow ^1S_0+^3D_1$ & $K K^*(1680)$ & $M^{JL}=M^{11}=-M^{000}$ \\
 & & $M^{000}=(\sqrt{\frac{1}{40}}I^{-1,-1}_{0,0}+\sqrt{\frac{1}{30}}I^{0,0}_{0,0}
    +\sqrt{\frac{1}{40}}I^{1,1}_{0,0})\phi_f$ \\ \hline
$X \rightarrow ^3S_1+^3P_1$ & $\rho a_1(1640), ~\rho a_1(1260), ~K^*K_1(1273),
 ~\omega f_1(1285)$  & $M^{JL}=M^{00}+M^{22}$ \\
 & & $M^{00}=\sqrt{\frac{1}{3}}(M^{0-11}-M^{000}+M^{01-1})$ \\
 & & $M^{22}=\sqrt{\frac{1}{6}}(M^{0-11}+2M^{000}+M^{01-1})$ \\
 & & $M^{0-11}=-\sqrt{\frac{1}{24}}(I^{0,0}_{0,0}+I^{1,1}_{0,0})\phi_f$ \\
 & & $M^{000}=\sqrt{\frac{1}{24}}(I^{-1,-1}_{0,0}+I^{1,1}_{0,0})\phi_f$ \\
 & & $M^{01-1}=-\sqrt{\frac{1}{24}}(I^{0,0}_{0,0}+I^{1,1}_{0,0})\phi_f$ \\ \hline
$X \rightarrow^3S_1+^3S_1$ & $\rho\rho, ~\rho\rho(1450),~\omega\omega, ~\omega\omega(1420),$
 & $M^{JL}=M^{11}=\sqrt{\frac{1}{2}}(M^{0-11}-M^{01-1})$ \\
 & $~K^*K^*, ~K^*K^*(1410),~\phi\phi$ & $M^{0-11}=\sqrt{\frac{1}{12}}(I^{0,0}_{0,0})\phi_f,
 ~M^{0-11}=-\sqrt{\frac{1}{12}}(I^{0,0}_{0,0})\phi_f$ \\ \hline
$X \rightarrow^3S_1+^1P_1$ & $\rho b_1(1235), ~K^*K_1(1400), ~\omega h_1(1170)$
 & $M^{JL}=M^{00}+M^{22}$ \\
 & & $M^{00}=\sqrt{\frac{1}{3}}(M^{0-11}-M^{000}+M^{01-1})$ \\
 & & $M^{22}=\sqrt{\frac{1}{6}}(M^{0-11}+2M^{000}+M^{01-1})$ \\
 & & $M^{0-11}=\sqrt{\frac{1}{12}}I^{1,1}_{0,0}\phi_f$, ~
     $M^{000}=-\sqrt{\frac{1}{12}}I^{1,1}_{0,0}\phi_f$ \\
 & & $M^{01-1}=\sqrt{\frac{1}{12}}I^{-1,-1}_{0,0}\phi_f$ \\ \hline
$X \rightarrow^3S_1+^3P_2$ & $\rho a_2(1320), ~K^*K^*_2(1430)$
 & $M^{JL}=M^{22}=-\sqrt{\frac{1}{2}}(M^{0-11}-M^{01-1})$ \\
 & & $M^{0-11}=\sqrt{\frac{1}{24}}(I^{0,0}_{0,0}-I^{1,1}_{0,0})\phi_f$ \\
 & & $M^{01-1}=\sqrt{\frac{1}{24}}(I^{-1,-1}_{0,0}-I^{1,1}_{0,0})\phi_f$ \\
\bottomrule
\end{tabular*}
\end{center}

\begin{multicols}{2}

All the possible decay modes of $\eta(\eta^{\prime})$ family are shown
in Table \ref{tab1}. To calculate the strong decay widths of mesons,
the strength of the
quark pair creation from the vacuum, $\gamma$, has to be fixed. It is
obtained by fitting the experimental values of the strong decay widths
of light and charmed mesons, charmonium and baryons. In the
present work, $\gamma=6.95$, which is adopted by many
researches~\cite{zhushilin,Blundell,yang}, is taken for the non-strange
quark pair creation, and the strength of
$s\bar{s}$ creation satisfies $\gamma_s$=$\gamma$/3~\cite{Yaouanc}.

\subsection{$\eta'(2^1S_0)$}
The experimental evidence for $\eta(1760)$ is controversial.
There are large differences between the observations of
MARK III, DM2 and BES
collaborations\cite{BESII,MARKIII1,MARKIII2,DM21,DM22}. In our calculation,
the mass of $\eta'(2^1S_1)$ is 1795 MeV, which is close to the
experimental mass of $\eta(1760)$. So we take it as the candidate of
$\eta(1760)$. In Fig.\ref{fig2}, we show the dependence of the partial widths
of the strong decay of the $\eta'(2^1S_0)$ on the $R_A$. Taking
$R_A$ =3.0-4.3 GeV$^{-1}$ discussed above, the total width ranges
from 256 to 404 MeV, which is much larger than the results given by Mark III and DM2
collaboration, but falls in the range of the BES experimental data.
In this range, $\eta'(2^1S_0)$ have a sizable branching ration into
$\pi a_0(980)$, $\pi a_2(1320)$, $\rho\rho$, and $KK^*$. But the
partial width to $\omega\omega$ is rather small. If the BES results
are reliable, the assignment of $\eta(1760)$ to $\eta'(2^1S_0)$ is disfavored
in the present calculation. In Ref.~\cite{PRD83}, $\eta(1760)$ is taken as
$\eta(3S)$, the total decay width is between 60-100 MeV, which falls in
the range of DM2's results, but is far below BES's results.
\begin{center}
\includegraphics[width=8cm]{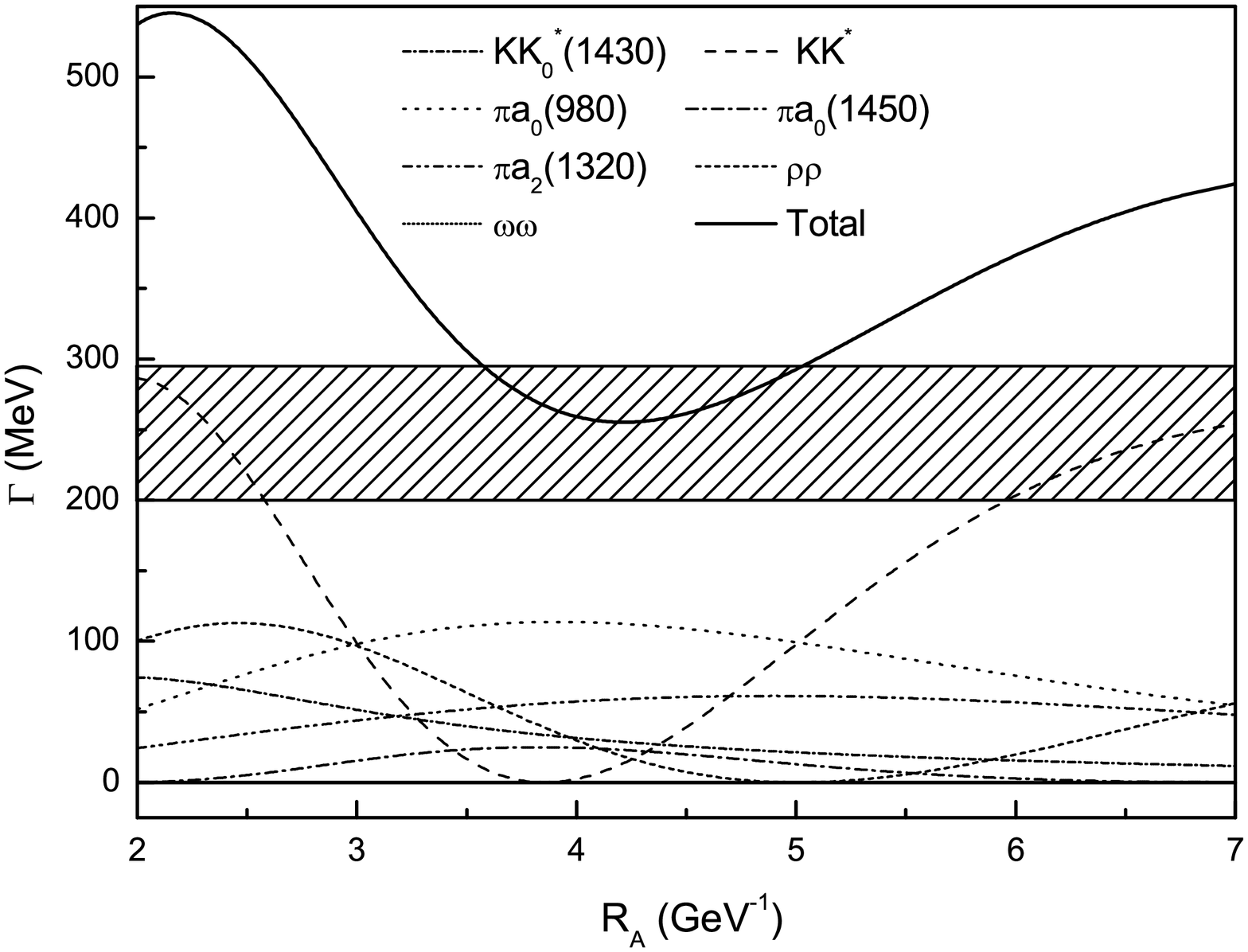}
\figcaption{The possible strong decay of the $\eta'(2^1S_0)$\label{fig2}}
\end{center}

\subsection{$\eta(4^1S_0)$}
$X(1835)$ was first observed by BESII in the $\pi^+\pi^-\eta'$
invariant-mass spectrum in the decay channel
$J/\psi\rightarrow\gamma\pi^+\pi^-\eta'$ with a statistical
significance of 7.7~$\sigma$\cite{BESII}. BESIII confirmed it in the
same process with statistical significance larger than
20$\sigma$~\cite{BESIII}. In the present calculation, the mass of
$\eta(4^1S_0)$=1807 MeV is close to the mass of $X(1835)$, so the
assignment of $X(1835)$ to $\eta(4^1S_0)$ is possible, which is different
from the assignment of Ref.~\cite{PRD83}, $\eta^{\prime}(3S)$.
In Fig.~\ref{fig2}, the dependence of the partial widths of the strong
decay of the $\eta(4^1S_0)$ on the $R_A$ is shown. From the mass
calculation, $R_A$ =4.6-7.5 GeV$^{-1}$ is obtained. In this range,
the total width ranges from 54 to 692 MeV, which
falls in the range of the BES experimental data, and the main decay
modes are $\pi a_0(980)$ and $\pi a_0(1450)$. We suggest experimental search for
$X(1835)$ in these modes to make sure whether it is $\eta(4^1S_0)$
assignment.
\begin{center}
\includegraphics[width=8cm]{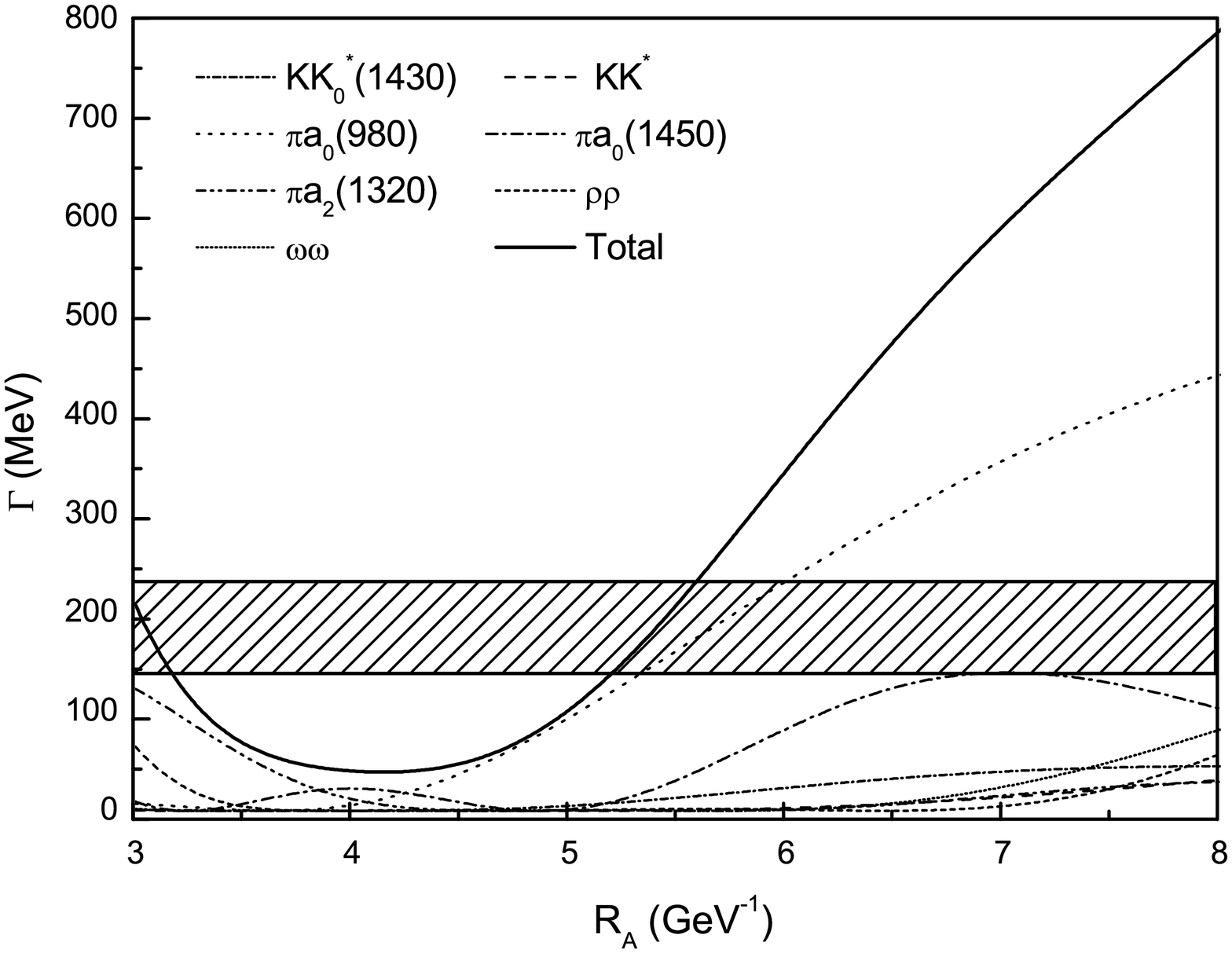}
\figcaption{The possible strong decay of the $\eta(4^1S_0)$\label{fig3}}
\end{center}

\end{multicols}
\ruleup
\begin{center}
\includegraphics[width=16cm]{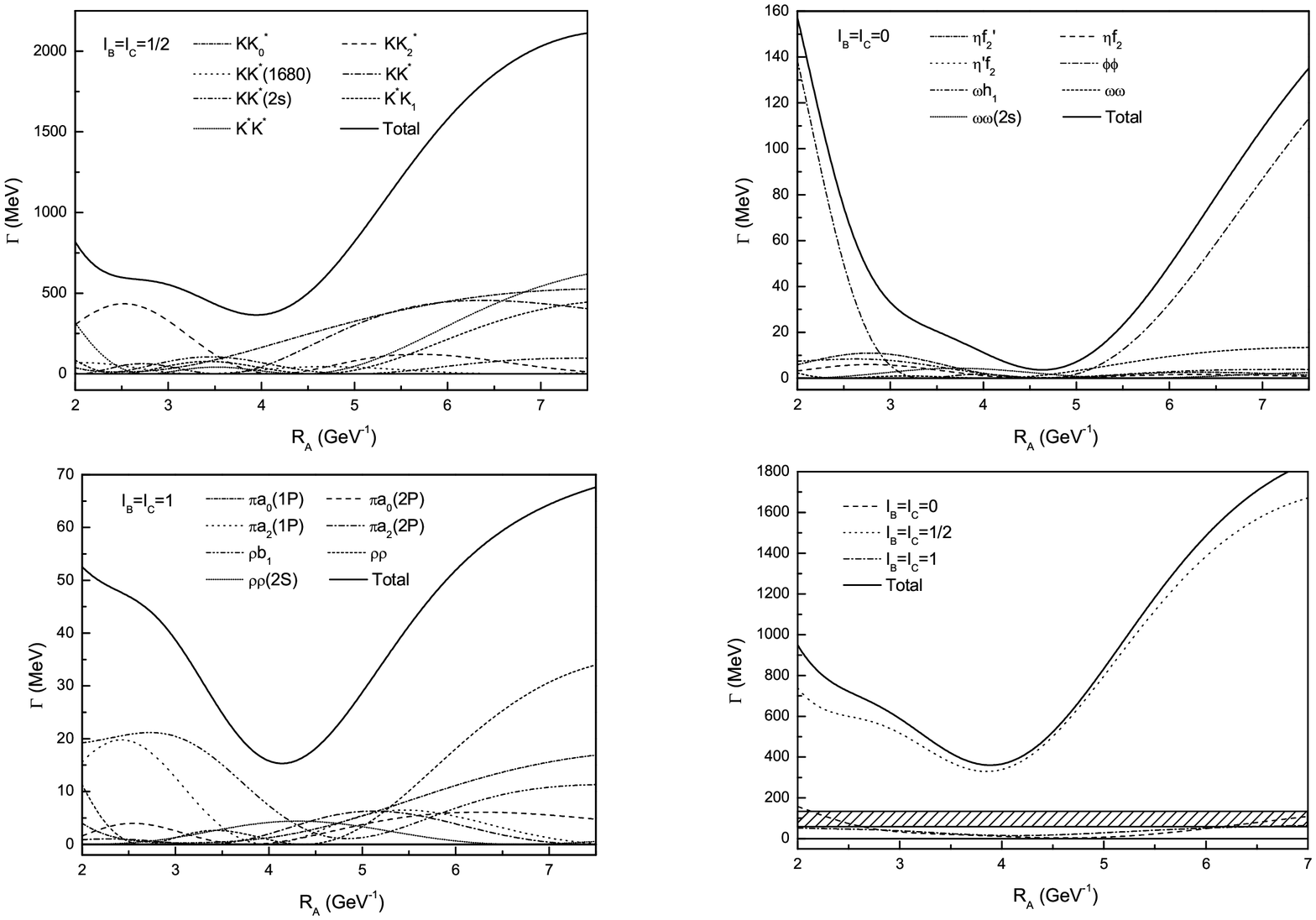}
\figcaption{The possible strong decay of the $\eta'(3^1S_0)$\label{fig4}}
\end{center}
\ruledown

\begin{multicols}{2}

\subsection{$\eta'(3^1S_0)$ and $\eta'(4^1S_0)$}

Besides confirmed the existence of $X(1835)$ in the $\pi^+\pi^-\eta'$
invariant-mass spectrum in the process $J/\psi \rightarrow \eta'\pi^+\pi^-$,
other two states $X(2120)$ and $X(2370)$ are observed by BESIII with
statistical significance larger than 7.2$\sigma$ and 6.4 $\sigma$,
respectively. By comparing the masses of the $\eta(\eta^{\prime})$ family,
it is possible to take $\eta'(3^1S_0)$ and $\eta'(4^1S_0)$ as the candidates of
$X(2120)$ and $X(2370)$. Because of their large masses, many strong decays
modes are allowed. In Figs.~\ref{fig4} and \ref{fig5}, the partial widths of
their strong decays are shown. For $\eta'(3^1S_0)$ with
$R_A$ =3.7-5.6 GeV$^{-1}$ and for $\eta'(4^1S_0)$ with $R_A$ =4.6-7.5 GeV$^{-1}$,
the decay widths are much higher than the experimental data of BESIII.
Because both $X_n$ and $X_s$ have contributions to the state $n\bar{s}s\bar{n}$,
the partial width of the strong decay to two isospin I=$\frac{1}{2}$ mesons is
generally much larger that to two isospin 1 or 0 mesons.
$\eta'(3^1S_0)$ have large partial to $KK^*_0$ and $KK^*$.
And the main decay modes of $\eta'(4^1S_0)$ are $KK^*$, $KK_1(1352)$,
$KK^*_0(1430)$, $KK^*_0(1950)$.

  If we describe $X(2120)$ and $X(2370)$ as $\eta'(3^1S_0)$ and
  $\eta'(4^1S_0)$ respectively with parameters in this work , it is not appropriate
  obviously.

\end{multicols}

\ruleup
\begin{center}
\includegraphics[width=16cm]{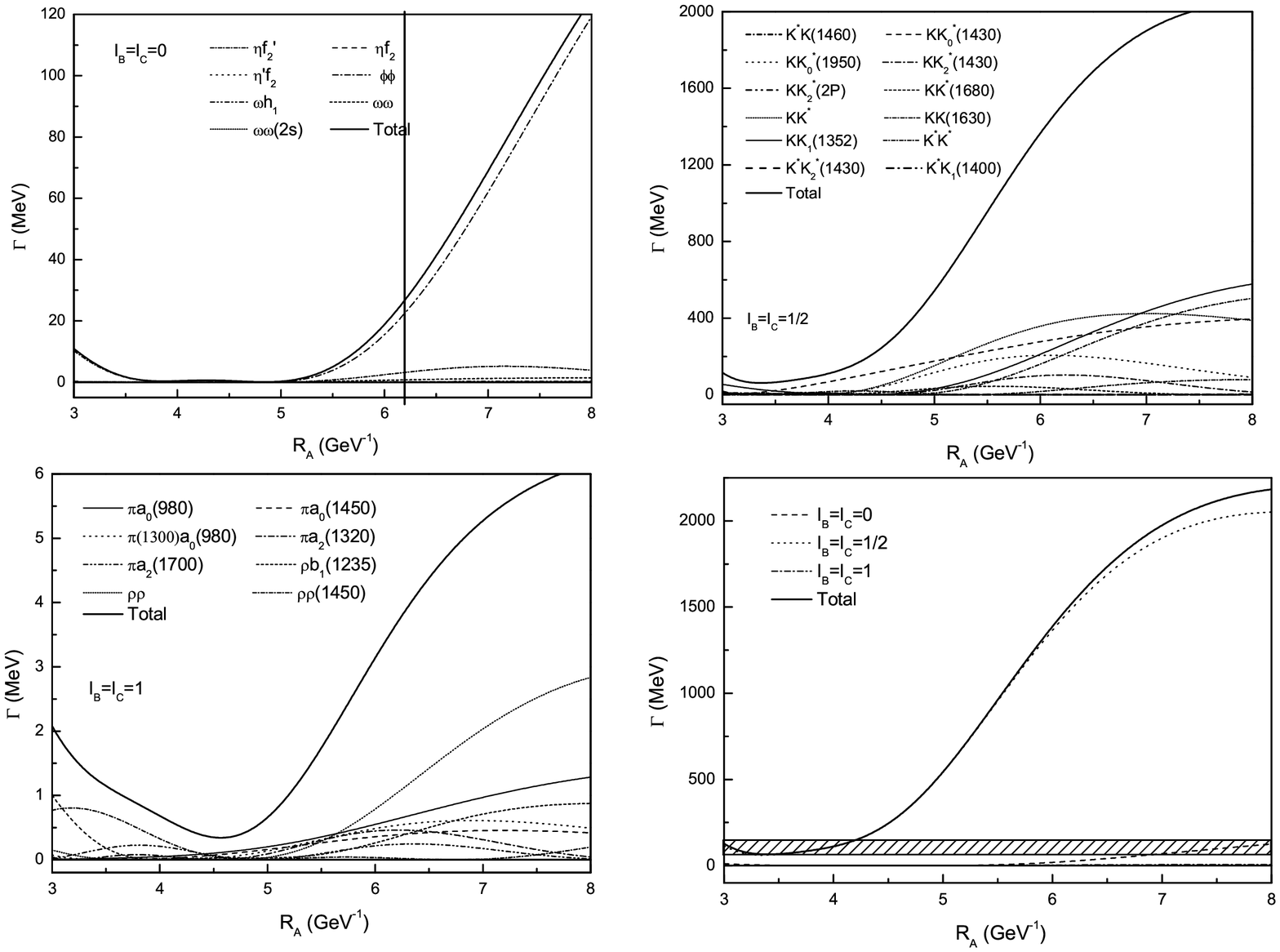}
\figcaption{The possible strong decay of the $\eta'(4^1S_0)$\label{fig5}}
\end{center}

\ruledown

\begin{multicols}{2}

\section{summary and discussions}
By using chiral quark model, the mass spectrum of $\eta(\eta^{\prime})$ family
are calculated, where the mixing between $(u\bar{u}+d\bar{d})/\sqrt(2)$ and
$s\bar{s}$ is determined by system dynamics, $K$-meson exchange.
Based on the mass spectrum, the possible candidates of
four $J^{PC}I^G=0^{-+}0^+$ mesons, $\eta(1760)$, $X(1835)$, $X(2120)$ and
$X(2370)$ are assigned to $\eta'(2^1S_0)$, $\eta(4^1S_0)$, $\eta^{\prime}(3^1S_0)$,
$\eta^{\prime}(4^1S_0)$. Furthermore, all kinematically allowed two-body strong
decays of them can calculated in the framework of the $^3P_0$ model.
The wavefunctions needed in the calculation are obtained from the mass calculation.
To simplify the calculation, the SHO wavefunctions are used to mimic the real
wavefunctions.

The decay widths turn out to be strongly dependent on the SHO wave function scale
parameter $\beta$. For $\eta(1760)$, the width is larger than the result of
\cite{DM22} and is compatible with the results of BES observation~\cite{BESII}
in the $R_A$ range. However the partial width to $\omega\omega$ is too small,
which it is incompatible with experimental date~\cite{MARKIII1,DM22,BESII}.
So the assignment of $eta(1760)$ to $\eta(2S)$ is disfavored in the present
calculation. For the state $X(1835)$, the calculated decay width is consistent
with experiment data, and $\pi a_0(980)$ and $\pi a_0(1450)$ are the main
decay modes. To justify this assignment, the experimental investigation of
the $\pi a_0(980)$ and $\pi a_0(1450)$ decay modes of $X(1835)$ is needed.
Sine $X(1835)$ is around the threshold of $p\bar{p}$, it may be the the
mixture of $q\bar{q}$ and baryonium. Further study of the state $X(1835)$
by taking into account of the mixture is essential to understand the
nature of the state.

$X(2120)$ and $X(2370)$ are assigned to $\eta^{\prime}(3^1S_0)$ and
$\eta^{\prime}(4^1S_0)$ respectively. Since they have larger masses,
many strong decays modes are allowed and have a large phase space
to some modes. The total decay widths are much higher than the
experimental values. The large decay width may de due to the
overestimated value of $\gamma$. To exclude the impact of parameters,
the branching ratio is better to justify the assignment. More experimental
data are needed. Since the lattice QCD predicts the
$0^{-+}$ glueball is about 2.3$\sim$2.6 GeV, which is around the
masses of $X(2120)$ and $X(2370)$, the study with the mixture of
$q\bar{q}$, glueball and other configurations are necessary to
understand the nature of $X(2120)$ and $X(2370)$ states.

\end{multicols}

\end{document}